\newcommand{\BE}{\begin{equation}}
\newcommand{\EE}{\end{equation}}
\newcommand{\BA}{\begin{eqnarray}}
\newcommand{\EA}{\end{eqnarray}}
\newcommand{\bx}{{\bf x}}

\newcommand{\bk}{{\bf k}}

\newcommand{\bxi}{{\boldsymbol\xi}}

\documentclass[graybox]{svmult}

\usepackage{type1cm}        
\usepackage{makeidx}         
\usepackage{graphicx}        
\usepackage{multicol}        
\usepackage[bottom]{footmisc}

\usepackage{newtxtext}       %
\usepackage[varvw]{newtxmath}       

\makeindex             

\begin{document}

\title*{Spatial competition and repulsion: pattern
formation and the role of movement 
}


 \author{Crist\'obal L\'opez, Eduardo H. Colombo, Emilio Hern\'andez-Garc\'\i a
 and  Ricardo Martinez-Garcia}

\institute{Crist\'obal L\'opez \at Instituto de F\'\i sica Interdisciplinar y Sistemas Complejos (IFISC), CSIC-UIB,
Campus Universitat Illes Balears, 07122 Palma de Mallorca, Spain \email{clopez@ifisc.uib-csic.es}
\and Eduardo H. Colombo \at Center for Advanced Systems Understanding, 
(CASUS) -- Helmholtz-Zentrum Dresden-Rossendorf (HZDR),
Untermarkt 20, G\"orlitz, 02826,
Germany 
\and Emilio Hern\'andez-Garc\'\i a \at Instituto de F\'\i sica Interdisciplinar y Sistemas Complejos (IFISC), CSIC-UIB,
Campus Universitat Illes Balears, 07122 Palma de Mallorca, Spain 
\and Ricardo Martinez-Garcia \at Center for Advanced Systems Understanding,
(CASUS) -- Helmholtz-Zentrum Dresden-Rossendorf (HZDR),
 Untermarkt 20, G\"orlitz, 02826,
Germany \at ICTP South American Institute for Fundamental Research \& Instituto de F\'isica Te\'orica, Universidade
Estadual Paulista - UNESP, R. Dr. Bento Teobaldo Ferraz, 271 - 2 - Várzea da Barra Funda, São Paulo - SP, 01140-070, Brazil 
}
%
%
\maketitle


\abstract{
This chapter investigates some mechanisms behind
pattern formation driven by competitive-only 
or repelling interactions, and
explores how these patterns are influenced by different types
of particle movement. 
Despite competition and repulsion
are both anti-crowding interactions, collective
 effects may lead to clusters of individuals,
which can arrange periodically.
Through the analysis of two models, it
provides insights into the 
similarities and differences in the
patterns formed and underlines
 the role of movement in shaping
the spatial distribution 
of biological populations. }

\section{Introduction}
\label{sec:intro}

Collections of interacting particles are widely used to study
complex systems, as they provide insights into diverse
collective behaviors observed in various fields, such as the
flocking of animals \cite{Cavagna2014}, spatial patterns in
ecosystems, embryos and colloids
\cite{Likos2007,Rietkerk2008,Baker2008,Borgogno2009}, cell
migration \cite{Alert2020}, and phase transitions in active
matter \cite{Cates2015}. Interactions among particles in these
models occur via forces and/or dynamical rules which can be
of physical, biological, or social origins, and  can
operate either locally, involving a few individuals, or over a
finite distance, influencing many individuals and leading to
long-range aggregates \cite{Murray2003,MG2022}.

The formation of these particle aggregates or clusters, that
may appear forming  global periodically ordered patterns, is
particularly significant for biological populations like
vegetation systems and cell populations \cite{Murray2003,MG2022}. The
spatial self-organisation has important ecological
implications, producing feedbacks with biological
functionalities that are fundamental for the biodiversity
maintenance,
population survival and  the
interactions with environmental
factors \cite{Wiegang2021}.
 A complete knowledge of the
conditions that lead to these spatial structures and their
feedback on the system dynamics is thus essential for the
understanding of complex living systems.

In recent years, the authors have explored different models of
pattern formation driven only by anti-crowding interactions
through repelling forces and intraspecific competition
(throughout this chapter we will refer to the both types of
interactions as competing), and examined how these patterns
are affected by particles' movement. Traditionally patterns in
natural systems have been interpreted in terms of variations of
the Turing mechanism \cite{Turing1952},
involving a combination of short-range
attraction (facilitation) to 
start aggregation, and long-range
repulsion (competition) to limit it spatially
\cite{Murray2003}. However, the pioneer works by Likos and
colleagues demonstrated that
repulsive-only interactions in colloidal solutions and polymers
could form cluster crystals, where clusters of particles
organize into a crystalline structure
\cite{Likos2007,Mladek2006}. In biological contexts, it has
been shown that competitive-only mechanisms (spatially
non-local) can give rise to vegetation pattern formation
\cite{Ricardo2013}, and species clustering in ecological niche
space can occur despite competitive exclusion
\cite{Scheffer2006,Pigolotti2007}.

This chapter presents results and discussions on two phenomena:
pattern formation through anti-crowding
 interactions (which prevent individuals from
 getting too close to each other), and the 
role of movement on the spatial structure.
 Two different models of motile
interacting individuals are discussed: one where particles repel
each other via forces deriving from
a soft potential, and another based on a
birth-death model with competition rules. These models are
studied using two complementary approaches: i) describing the
particle dynamics, mainly based on off-lattice numerical
simulations, and ii) a continuum mathematical description using
partial differential equations for the number
density of particles. Despite their differences, both models
form the same type of patterns (cluster crystals) and, most
importantly, under similar properties for the interaction.
However, the underlying physical and/or biological mechanisms
are completely different. The chapter also examines the
influence of the type of movement of the organisms on the
spatial structures. It has been noted that movement patterns of
some living organisms are consistent with L\'evy flight
behavior \cite{Metzler2000,Dieterich2008,Viswanathan2008},
which has been argued to be advantageous over standard Brownian
motion in certain foraging strategies \cite{Bartumeus2003}.
This advantage primarily stems from the occurrence of
occasional long jumps. These findings (and many others)
highlight the importance of the type of movement when studying
the spatial self-organization of biological populations.
However, the impact of L\'evy-type diffusion on the properties
of organism aggregates has not been extensively studied.
Moreover, the importance of self-propelled  motion (by which
particles move consuming internal energy) is key to understand
living systems \cite{Marchetti2013}. We will briefly present
some results concerning cluster formation with active
(repelling) and L\'evy particles.

The structure of this chapter is as follows. In Section
\ref{sec:models} we present a comparative study of the pattern
formation dynamics in the two discrete particle models. In
Section \ref{sec:motion} we analyze the impact of the type of
motion on the spatial structures. Finally, Section
\ref{sec:summary} provides a summary and discussion.

\section{Mechanisms of pattern formation
for repelling/competing individuals}
\label{sec:models}

Here we present a comparative study of the pattern
formation dynamics in two types of interacting particle
systems. The first system consists of Brownian particles
interacting via a repulsive soft potential \cite{Delfau2016}.
The second one is a set of Brownianly moving particles, with no
forces among them but with a birth-death dynamics
\cite{EHG2004,Lopez2004}. For both models,
we provide their mean-field density
equation description to highlight the similarities in the
interaction and diffusion conditions required to obtain spatial
periodic patterns, and discuss the different physical
mechanisms leading to these patterns.

\subsection{A system of moving individuals with repulsive forces}
\label{subsec:1}

\begin{figure}[b]
\sidecaption
\includegraphics[scale=.4]{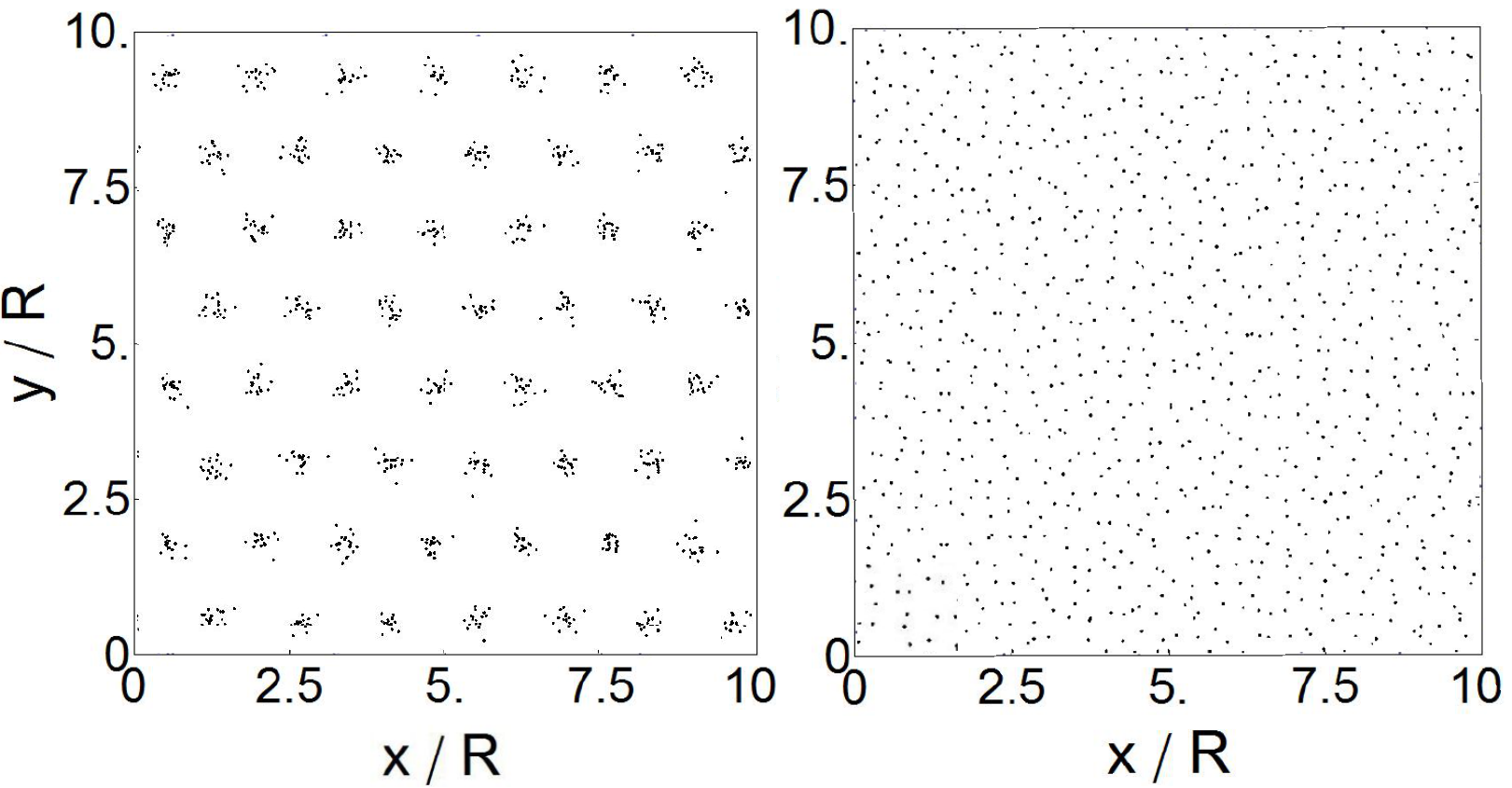}
%
%
\caption{
Snapshot of the positions of $N=1000$ particles at large times, moving according to
Eq. (\ref{Brownian}),
for $R=0.1, \epsilon=0.033$,
and $L=1$. Left plot is for the GEM-3 potential, i.e, $\alpha=3$
and $D=0.02$. The right plot is for GEM-1  ($\alpha=1$) and $D=0.005$.
Modified from \cite{Delfau2016}.
}
\label{fig:1}       
\end{figure}

Let us consider a system of $N$ Brownian particles in contact
with a thermal bath, in the overdamped limit, contained in a
periodic two-dimensional domain. They interact through a
soft-core (i.e. a potential not divergent at ${\bf
x}=0$) repulsive two-body potential. The motion of the
particles is given by
\BE
\dot\bx_i = - \sum_{j=1}^N\nabla V(\bx_i-\bx_j)
+\sqrt{2D}~\bxi_i(t) \ ,
\label{Brownian}
\EE
where the diffusion coefficient is
$D=k_B T/\gamma$, with $\gamma$
the friction coefficient, $T$ the temperature of the bath,
 and
Gaussian noise vectors
$\bxi_i$ satisfying
\BE
\left<\bxi_i\right>=0 \ , \
\left<\bxi_i(t)\bxi_j(t')\right>=\mathbb{I}\delta_{ij}\delta(t-t')
\ .
\EE
For the interaction potential we consider the generalized
exponential model of exponent $\alpha$ (GEM-$\alpha$),
parameterized with exponent $\alpha$, widely used in many
applications \cite{Mladek2006,Pigolotti2007}:
\BE
V(\bx)=\epsilon
\exp\left(-\left|\frac{\bx}{R}\right|^\alpha\right)\ .
\label{GEM_def}
\EE
$\epsilon$ is an energy scale and $R$ a characteristic
interaction range. Numerical simulations (see Fig.~\ref{fig:1})
confirm that hexagonal patterns spontaneously appear if $\alpha
>2$ and the diffusivity is small enough. Patterns do not occur
if $\alpha \leq 2$, independently of the value of the diffusion
coefficient. For any $\alpha$, large values of the
diffusivities give rise to a homogenous spatial distribution
of particles.

This can be analytically understood by considering the
Dean-Kawasaki (DK) equation \cite{Dean1997} describing the
system dynamics for the microscopic density of particles
$\rho(\bx,t)=\sum_{i=1}^N \delta\left(\bx-\bx_i(t)\right)$. In
fact, to reproduce the phenomenology presented here it is
enough to consider its deterministic version, i.e. the one
obtained by neglecting stochastic noise terms, 
which is appropriate
 for sufficiently large density of particles, 
 and amounts to
a kind of mean-field description:
\BA
\partial_{{t}} \rho({\bx},{t}) = \nabla \cdot
\left({\rho}({\bx},{t})\int d{\bx}' \nabla
{V}({\bx}-{\bx'}){\rho}({\bx}',{t})\right)
+ {D}\nabla^2 \rho({\bx},{t}).
\label{Eq:Dean}
\EA

The diffusion term results from the random 
motion of Brownian particles, while the 
potential term describes density 
advection by the local velocity due to repulsive forces.
Note the conservative character of this equation (it is of
continuity type) reflecting that the total number of particles
in the system remains constant. Numerical simulations confirm
that it properly describes the density of the particle system
at large scales \cite{Delfau2016}, in regimes where
fluctuations can be neglected.
Any  constant density  $\rho_0$ is a solution of
Eq.~(\ref{Eq:Dean}). We perform a linear stability analysis
around this homogeneous solution by considering a small
perturbation $\rho(x,t) =  \rho_0 + \exp{(\lambda t + i
\boldsymbol{k} \cdot \boldsymbol{r})}$, which gives the
following dispersion relation, or growth rate:
\BE
\lambda (k) = -k^2 \left[ D + \rho_0 \hat V (k)\right],
\label{disp_nonoise}
\EE
with $\hat V (k)$  the Fourier transform of the interaction
potential ($k=|\bk|$).
A positive value of $\lambda (k)$ for some $k$ reveals a
pattern-forming instability in which perturbations of
periodicity $2\pi/k$ will grow on top of the homogeneous
density $\rho_0$, leading to pattern formation with that
periodicity. Note that
 a necessary condition for $\lambda
(k)>0$ is that the Fourier transform of the potential takes
negative values for some values of $k$ \cite{Lopez2004}. This
explains why patterns can never be observed with, for example,
GEM-1 or GEM-2 potentials, whose Fourier transforms are always
positive. In general, the Fourier transform of the GEM-$\alpha$
potential
is always positive if $\alpha \leq 2$. This is consistent with
numerical simulations of the density equation,
Eq.~(\ref{Eq:Dean}), which lead to a homogeneous state for
$\alpha \le 2$ \cite{Delfau2016}, and analogously for the
particle system. For $\alpha >2$, however, the potential is
more box-shaped, leading to Fourier transforms that may have
negative values \cite{Pigolotti2007}. Note that for the
derivation of Eq. (\ref{disp_nonoise}) to hold, the soft
character of the potential is crucial since otherwise, i.e.,
the hard-core case, its Fourier transform is not properly
defined and another methodology is needed \cite{Caprini2018}.
Also, in Eq.~(\ref{disp_nonoise}) we observe that 
 the growth rate $\lambda (k)$ is always negative
for large values of $D$, so that
large diffusion coefficient 
inhibits pattern formation and
makes the spatial distribution 
of particles homogeneous.

\begin{figure}[b]
\sidecaption
\includegraphics[scale=.65]{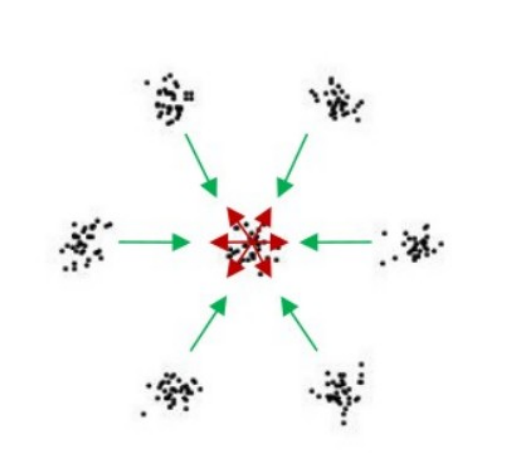}
\caption{
Schematic representation of balance of
forces between neighboring clusters (green)
and neighboring particles inside a cluster (red).
}
\label{fig:2}       
\end{figure}

Eq.~(\ref{disp_nonoise}) shows the mathematical conditions
for the instability of the homogenous solution in the continuum
density description of the system of particles. A detailed
study performed in \cite{Delfau2016} indicates that physically
the formation of clusters comes from a balance between the
internal repulsion of particles inside a cluster, and the
external repulsion from the particles in neighboring clusters
(see Fig. \ref{fig:2}), and mediated by the diffusion of the
particles. Thus, when diffusion is small and repulsion from
particles in neighboring clusters dominates the internal
repulsion with other particles in the same cluster, aggregation
tends to be enhanced.

\subsection{A system
of birth-death particles}
\label{subsec:2}

Let us consider a system of initially $N_0$ particles
performing random Brownian motion with diffusivity $D$ in a
two-dimensional square with periodic boundary conditions. In
addition, the particles undergo a birth-death dynamics, so that
the total number of them is not constant, and there are $N(t)$
particles in the system at time $t$. The demographic events
occur at stochastic Poisson times with the following rates per
particle: i) at rate $\beta_0$ one particle is chosen to die,
i.e. to disappear from the system; ii) particle $i$ is chosen
with rate $\lambda_i$ to reproduce, with the offspring being
placed at the same location as the parent. The important point
is that the birth rate $\lambda_i$ diminishes with the number
of particles within a radius $R$ from the focal particle $i$.
This is a kind of competing dynamics since the birth rate of
a given individual decreases with the number of others in its
surroundings, which maybe due to competition for resources or
mates. More explicitly, the  rate $\lambda_i$ at which
a new particle is introduced in the system, exactly at the
location of the parent particle $i$, is $\lambda_i= \lambda_0 -
g N_R^i$, where $\lambda_0$ and $g$ are constants, and $N_R^i$
is the number of particles around $i$ within a radius $R$.

We can consider more generally that this neighborhood
dependence is weighted with the distance. The weight is
expressed through an {\it influence} function that
decays with the
distance between the particles and the focal one
  $G(\bx_{i}-\bx_j)$, $j=1,...,N(t)$.
In this way, particle $i$ will give rise to the birth of a new
particle at its position with a probability rate
$\lambda_i=\lambda_0-g\sum_{j\neq i} G(\bx_i-\bx_j)$. The
previous case is recovered for a top-hat function $G(\bx)$,
i.e. a constant if $|\bx|<R$ and 0 otherwise.

\begin{figure}[b]
\includegraphics[scale=.4]{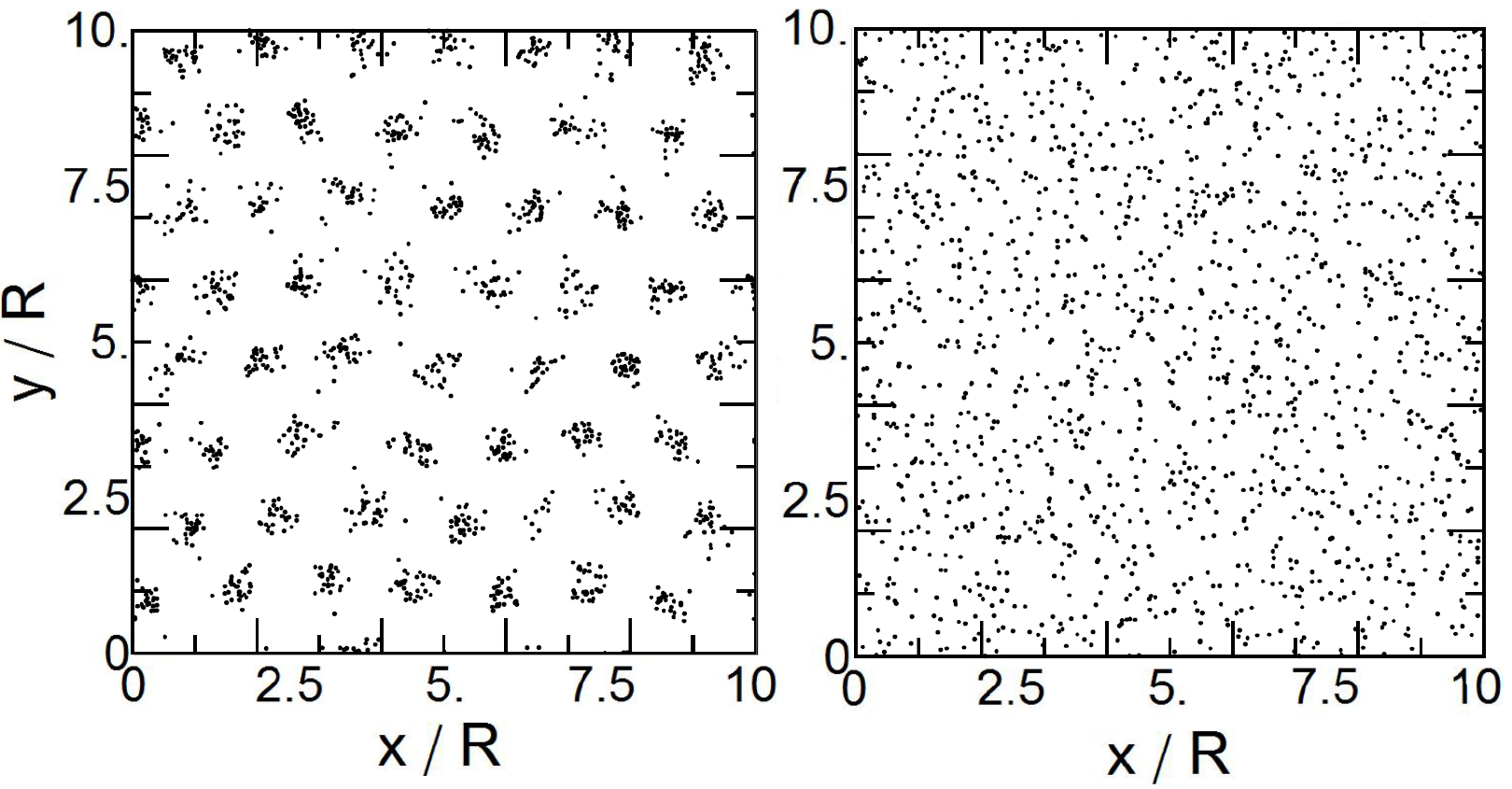}
\caption{Long-time spatial structures for the birth-death model
with the top-hat influence function, i.e. birth rate $\lambda_i=\lambda_0-g N^i_R$,
where $N^i_R$ is the number of neighbors within distance $R$ from particle $i$.
$R=0.1$, system size $L=1$, $g=0.02$. Left: $D=10^{-5}$, $\lambda_0=0.85$ and $\beta_0=0.15$,
so that $\mu=\lambda_0-\beta_0=0.7$. Right: $D=10^{-4}$, $\lambda_0=0.95$ and $\beta_0=0.05$,
so that $\mu=0.9$. The largest diffusion coefficient destroys the spatial pattern.
Modified from \cite{EHG2004}.
}
\label{fig:3}       
\end{figure}

Two transitions can be identified in the system:
\begin{enumerate}
\item An absorbing transition that differentiates between
    the absorbing phase, where all particles are dead (i.e.
    there are no particles in the system), and an active
    phase. In the active phase, at large times and high
    values of $\mu =\lambda_0 -
    \beta_0$ (the parameter driving the transition), the
    system contains many particles.

\item Depending on the shape of the influence function,
    $G$, and if the diffusion coefficient of the particles
    is sufficiently small, the particles may arrange into a
    hexagonal pattern, as illustrated in Fig. \ref{fig:3}
    for a top-hat influence function. The plot also shows
    the statistically homogeneous distribution that occurs
    when the diffusion coefficient is large.
\end{enumerate}

Note that the pattern structure resembles that observed in the
system of repulsive particles (Fig. \ref{fig:1}). However there
are significant differences  between the two models. The system
of repelling particles is in equilibrium with a thermal bath
and the total number of particles is conserved. In contrast,
the birth-death system is a prototypical non-equilibrium
system, characterized by the presence of an absorbing
configuration, and the number of particles fluctuates and is
not conserved (although its average reaches a constant value at
long times).

\begin{figure}[b]
\sidecaption
\includegraphics[scale=.65]{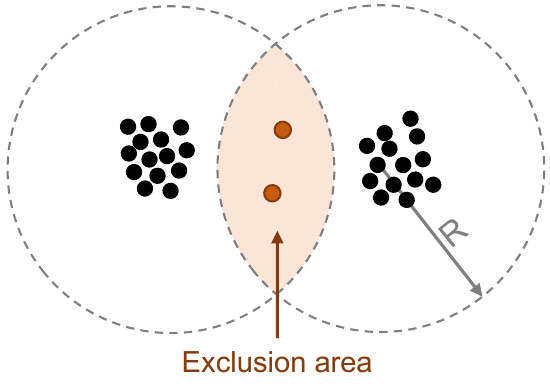}
\caption{Schematic representation of the formation of
exclusion areas (orange shaded region), where
individuals
have to compete with two
different clusters, whereas individuals in each patch
compete only with individuals in its own patch.
}
\label{fig:4}       
\end{figure}

We can further explore the analogies and differences by
examining the deterministic macroscopic density equation for
this system  with an arbitrary influence function $G$. This
equation was derived using field theoretical techniques in
\cite{EHG2004}. As before, we neglect here the stochastic terms
so that a mean-field version of it is given by
\BE
\partial_t \rho(\bx,t) =  \rho(\bx,t)
\left(\mu - g \int d\bx' G(\bx-\bx') \rho(\bx',t) \right) + D
\nabla^2 \rho(\bx,t). \label{BD_nondim}
\EE
Here, the influence function is usually normalized so that
$\int G(\bx) d\bx =1$. Note that Eq. (\ref{BD_nondim}) is not a
continuity equation as Eq.~(\ref{Eq:Dean}), indicating that the
total number of particles is not conserved now. A linear
stability analysis of Eq.~(\ref{BD_nondim}) is also
straightforward. Again considering a perturbation around the
homogenous solution, $\rho_0=\mu/g$, we obtain  the following
growth rate
\BE
\lambda (k) = -(D k^2 +\mu \hat G(k)), \label{BD:growthrate}
\EE
where $\hat G(k)$ is the Fourier transform of the influence
function. Note the similarity between this expression and
Eq.~(\ref{disp_nonoise}). A necessary condition for pattern
formation is that the Fourier transform of the influence
function has negative values, which is the same condition
required for the potential in the system of repelling particles
(numerical simulations of the density equation,
Eq.~(\ref{BD_nondim}), confirm this scenario \cite{EHG2004}).
But the underlying mechanism is quite different. Patterns
appear due to the presence of exclusion zones, which are
spatial zones between the clusters (at distances larger than
$R$ and smaller than $2R$) that fall within the range of
influence of two clusters. In these zones, particles face
enhanced competition  as they must compete with particles from
both clusters. This phenomenon is similar to that observed  in
some nonlocal vegetation models, where exclusion areas were
 first identified \cite{Ricardo2013}.
See Fig. \ref{fig:4} for a visual explanation of this concept.
Note that, when spacing is fixed
by the interaction range, 
hexagonal patterns naturally result from 
self-organization due to competitive 
interactions. 
This configuration maximizes
the exclusion area 
while maintaining 
spacing. However, more
complex models
can produce other structures
like stripes \cite{Ricardo2014}.

\subsection{Discussion on the influence function}
\label{sec:influence}

The models discussed in the previous section are
phenomenological effective descriptions of populations of
interacting individuals. Real interactions, specially in
biological contexts, are often mediated by agents, visual or
chemical signals, or substances (in general called mediators)
whose temporal dynamics is usually faster than that of the
population. A more fundamental description of the system should
explicitly consider these signal dynamics, with the effective
mathematical equations (e.g. Eq.~(\ref{BD_nondim})) derived
through the adiabatical elimination of the signal dynamics.
However, there is no systematic derivation of the population
dynamics leading to arbitrary influence functions keeping the
character of competitive-only interactions. In cases where such
a derivation has been provided, the resulting equations did not
have the characteristics necessary for pattern formation
\cite{Ricardo2014} (in vegetation systems), or, although
leading to patterns, the influence functions lack of generality
\cite{Mogilner1995} (in the context of cell dynamics).

A new mechanism has been proposed that allows for the
derivation of more general influence functions, including those
with a shape similar to the GEM-$\alpha$ function
\cite{Eduardo2023,Eduardo2024}, in particular with cases
leading to non-positive Fourier transforms. Briefly, this
mechanism considers that mediators are released into the
environment in the form of short pulses. The response to these
pulse releases can significantly change the spatial stability
of the population. In particular, in the limit of fast mediator
dynamics, an effective description for the single population
emerges, where the influence function is general enough and
capable of leading to spatial instabilities.

\section{Role of motion on spatial structure}
\label{sec:motion}

In the models of the previous Section, 
despite their profound differences, we
observed that the same ingredients are key for pattern
formation: the spatial shape of the function
(the potential or the influence) giving the
particle interactions and a low enough diffusivity. Having
focused on the role of interactions in the earlier sections, we
will now discuss deeper the role of motility, considering two
types: L\'evy flights and active motion. First, we will analyse
the impact of L\'evy on the birth-death model, and then we will
briefly examine the effects of active motion on the system of
repulsive particles. Given the numerous analogies in the
descriptions of both models, many of the results discussed for
one system are applicable to the other.

As demonstrated both using the discrete particle dynamics and
the instability analysis of the continuum description, clusters
disappear and the population shows a uniform distribution for
Brownian particles moving {\it very randomly}, i.e., when their
diffusion coefficient is large. A natural question arises: what
happens in the system if particles exhibit other types of
motion? Of particular biological relevance is to consider
particles performing L\'evy flights or active motion.

L\'evy flights results in movement patterns that are
statistically different to Brownian motion. The displacement lengths
have a power-law distribution resulting in the alternation of
frequent, short displacements, with rare, long ones. Due to the
weight of these long jumps, the variance of the displacement
length is infinite, which translates into an infinite diffusion
coefficient. Yet, one can still define a generalized diffusion
coefficient, $\kappa_\gamma$, such that the  typical
displacement of an individual particle scales as $x \propto (\kappa_\gamma
t)^{1/\gamma}$, where $0<\gamma<2$ is the L\'evy index or
anomalous exponent which determines the type of motion. The
smaller the value of $\gamma$ the more anomalous is the random
walk (more frequent are long jumps). The question is whether
the formally infinite diffusion coefficient destroys the
spatial patterns. To analyze the impact of L\'evy type
of motion and spatial patterns we, as mentioned, will focus on
the birth-death system (see \cite{Heinsalu2010}). We consider
the same demographic processes as in Sec.\ref{subsec:2} but
instead of Brownian motion the particles perform L\'evy flights.

\begin{figure}[b]
\sidecaption
\includegraphics[scale=.23]{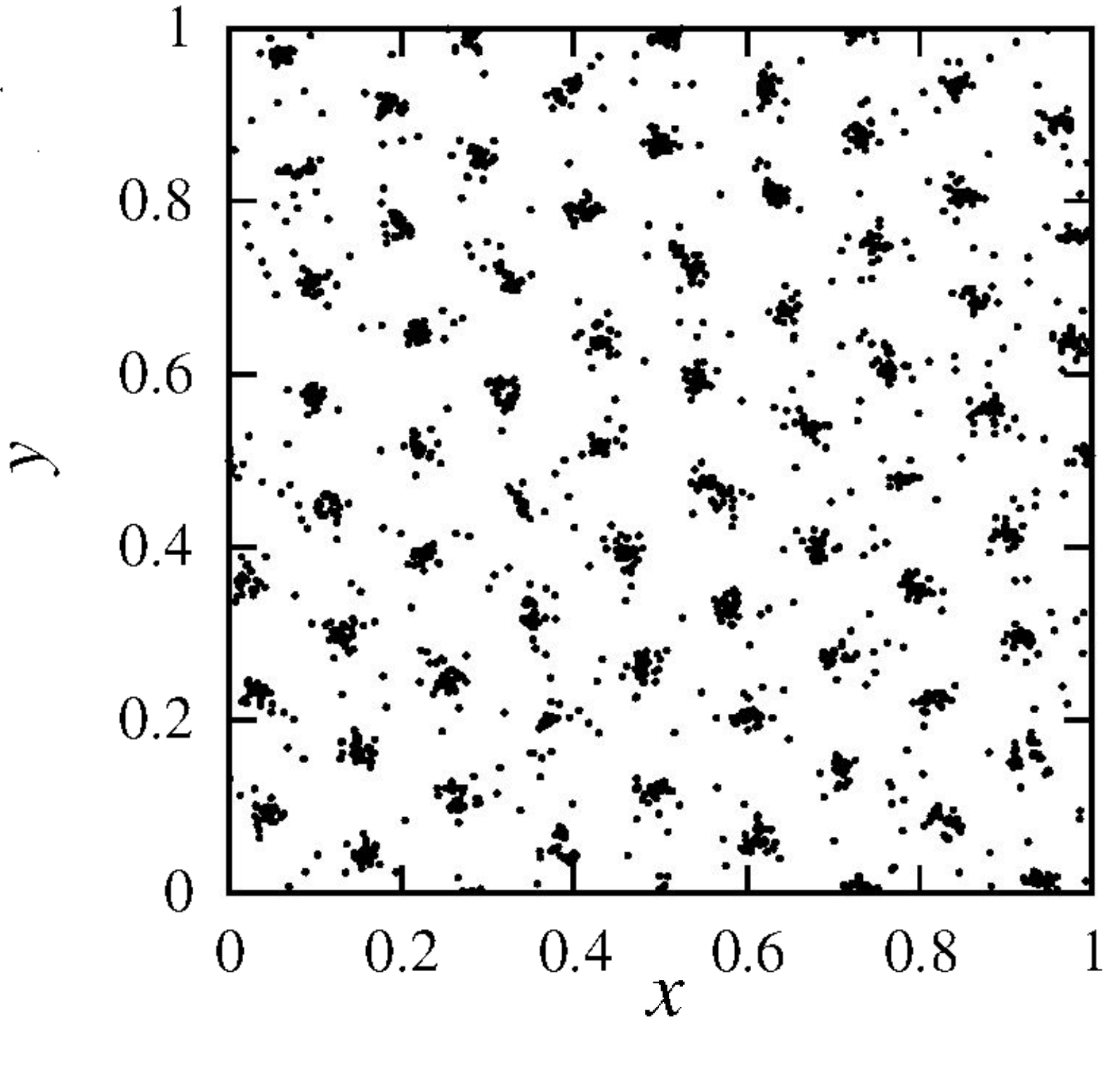}
\caption{
Particle configuration at long times for particles interacting as in Fig. \ref{fig:3},
but with L\'evy motion with L\'evy index $\gamma=1$ and $\kappa_\gamma=
56 \times 10^{-5}$. Other parameters: $R=0.1$, $L=1$, $\lambda_0=1$,
$\beta_0=0.1$, $g=0.02$.
Modified from \cite{Heinsalu2010}.
}
\label{fig:5}       
\end{figure}

As shown in \cite{Heinsalu2010}, see also Fig.~\ref{fig:5},
clusters are not broken by the infinite diffusion coefficient
of L\'evy flights. However, large jumps do influence the
characteristics of the patterns: clusters are less compact, and
many single particles are found moving between them. The
large-scale properties of the system can be derived from the
mean-field density equation, which now, in terms of the
fractional Laplacian $\nabla^\gamma$, takes the form:
\BA
\partial_{{t}} \rho({\bx},{t}) =  {\rho}({\bx},{t})
\left( \mu - g \int d{\bx}'
{G}({\bx}-{\bx'}){\rho}({\bx}',{t})\right) + \kappa_{\gamma}
\nabla^{\gamma} \rho({\bx},{t}).
 \label{Eq:levy}
\EA
Doing a stability analysis similar to the previous cases we
obtain the following perturbation growth rate
\BE
\lambda (k) = -(\kappa_\gamma k^\gamma +\mu \hat G(k)),
\label{Levy:growthrate}
\EE
where $\hat G(k)$ is the Fourier transform of the influence
function. There are no essential differences between
Eq.~(\ref{Levy:growthrate}) and Eq.~(\ref{BD:growthrate}). As
shown in \cite{Heinsalu2010} the L\'evy index $\gamma$ has only
a small impact on the properties of the pattern.

\begin{figure}[b]
\sidecaption
\includegraphics[scale=.35]{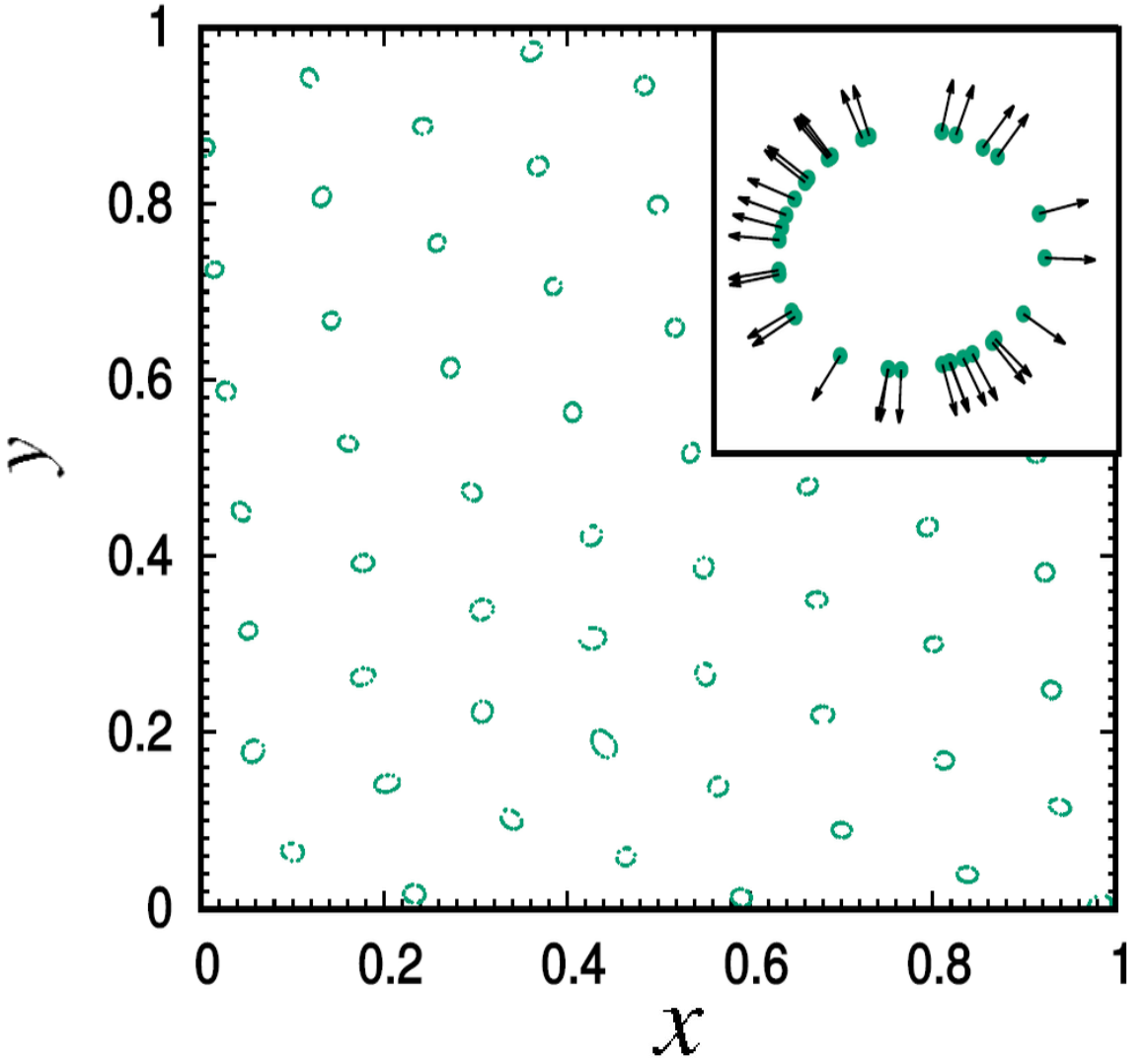}
\caption{
Ring-shaped clusters appearing in a system
of self-propelled repulsive particles.
The inset shows the orientation of the
self-propulsion velocity of the
particles in a given cluster.
Taken from \cite{Caprini2019}.
}
\label{fig:6}       
\end{figure}

One can conclude that the large-scale collective behavior of
the system is more strongly influenced by the competitive
interactions than by the type of motion performed by the
particles. The long flights characteristics of L\'evy movement
do not accumulate their effect enough to break clusters.
However, this is of course not always the case, as demonstrated
by the consideration of active motion.
\cite{Delfau2017,Caprini2019} explored a simple extension of
the system of repulsive particles
Eq.~(\ref{Brownian}) by introducing an internal degree of
freedom: the orientation of a self-propulsion speed. This
results in a system of active Brownian particles interacting in
pairs through a repulsive potential. Briefly, activity has two
main effects on the distribution of particles (when the
diffusivity of the particles is small, and thus clusters form
in the absence of activity): a) For high values of the
self-propulsion velocity, clusters are destroyed, and the
steady-state becomes statistically homogeneous (the cluster
crystal melts). This effect is  similar to an enhanced
effective diffusivity of the particles. b) For intermediate
values of the self-propulsion velocity, particle density within
clusters is depleted, see Fig.~\ref{fig:6}, so that they reach
a ring-like shape. This is a striking effect induced by the
self-propelled motion.

\section{Summary and conclusions}
\label{sec:summary}

We have presented a comparative study of two types of motile
competing particle systems. Despite their significant
differences, both of them exhibit similar spatial patterns that
are induced by the same conditions for the interactions and the
diffusivity of the individuals. However, the underlying
physical or biological mechanisms differ substantially. We also
analyzed the role of motion on patterns characteristics. In the
case of L\'evy flights, we concluded that  their impact is less
significant than  that of the interactions. Competitive
interactions are the main drivers of the global spatial
structure of the system. However, striking impacts were
observed when we considered active motion, for example
making single aggregates of particles become rings.

The  examples we showed highlight an important observation in
the literature on pattern formation studies: while analyzing
the properties of structures can be challenging, it often does
not reveal the underlying natural mechanisms or the
functionalities of the individuals that are organized in such
spatial distributions.
However, numerous studies show that pattern formation is key to
understand the factors leading to survival, the proper response
to external factors, and ecological and evolutionary processes
in biological dynamics. For instance, there is a known spatial
pattern sequence to desertification in water-limited
ecosystems, so that it may serve as
early warning signal \cite{Kefi2007}. 
At much smaller scales,
the spatial organization
of cells is essential for
 accurately characterizing,
forecasting and controlling 
tumor evolution \cite{Bellomo2008}.
For non-neutral species, 
patterns may enable  weaker competitors
to survive in conditions 
that would otherwise lead to their 
extinction when spatial 
distribution is ignored \cite{Maciel2021}.
In the context of the birth-death model we presented, it has
been shown that when two otherwise identical types of particles
compete, the type dispersing less and forming more compact
clusters, due to its type of motion, is more likely to survive
\cite{Heinsalu2013}.

Regarding the interplay between spatial dispersion of organisms
and their mobility, it is worth mentioning an emerging field
known as proliferating active matter, which considers
demographic dynamics in addition to the other factors. This
field combines elements of the two models presented in this
chapter, along with self-propulsion, and offers a natural
framework for studying many biological systems
\cite{Hallatschek2023}.

In conclusion, we believe that gaining a deeper understanding
of the feedbacks between spatial structure, interactions and
individual motility will provide valuable insights into the
dynamics of living complex systems.

\begin{acknowledgement}

C.L. acknowledges the warm hospitality at the Isaac Newton
Institute for Mathematical Sciences (INI) during the program
{\it Mathematics of movement: An interdisciplinary approach to
mutual challenges in animal ecology and cell biology.} 
He also acknowledges the Scultetus
Visiting Scientist Program at the
 Center
for Advanced Systems Understanding (CASUS)
in G\"orlitz, Germany, 
where this work was
completed.
C.L. and
E.H-G. acknowledge grant LAMARCA PID2021-123352OB-C32 funded by
MCIN/AEI/10.13039/501100011033323 and FEDER "Una manera de
hacer Europa"; and Grant TED2021-131836B-I00 funded by
MICIU/AEI/10.13039/501100011033 and by the European Union
"NextGenerationEU/PRTR". E.H-G. also acknowledges the Maria de
Maeztu program for Units of Excellence, CEX2021-001164-M funded
by MCIN/AEI/10.13039/501100011033.
E.H.C. and R.M-G. are partially 
funded by the Center of Advanced Systems
Understanding (CASUS), which is 
financed by Germany’s Federal Ministry
of Education and Research (BMBF) 
and by the Saxon Ministry for Science,
Culture and Tourism (SMWK) with 
tax funds on the basis of the budget
approved by the Saxon State Parliament.

\end{acknowledgement}
%
\eject
%
%

\end{document}